\documentclass[a4paper,fleqn]{cas-dc}
\usepackage{caption}
\usepackage{graphicx}
\usepackage{array}
\usepackage[sort, numbers]{natbib}
\setcitestyle{square}
\def\tsc#1{\csdef{#1}{\textsc{\lowercase{#1}}\xspace}}
\tsc{WGM}
\tsc{QE}
\tsc{EP}
\tsc{PMS}
\tsc{BEC}
\tsc{DE}

\begin{document}
\let\WriteBookmarks\relax
\def\floatpagepagefraction{1}
\def\textpagefraction{.001}
\shorttitle{Multi-Speaker Emotional Speech Synthesis}
\shortauthors{A Shahid et~al.}
\title [mode = title]{Generative Emotional AI for Speech Emotion Recognition:\\ The Case for Synthetic Emotional Speech Augmentation}



\author[1]{Abdullah Shahid}

\address[1]{Information Technology University (ITU),Punjab, Pakistan}

\author[2]{Siddique Latif}
\cormark[1]

\author[3]{Junaid Qadir}
\address[2]{University of Southern Queensland, Australia}

\address[3]{Qatar University, Doha, Qatar}

\cortext[cor2]{Corresponding author}

\begin{abstract}
Despite advances in deep learning, current state-of-the-art speech emotion recognition (SER) systems still have poor performance due to a lack of speech emotion datasets. This paper proposes augmenting SER systems with synthetic emotional speech generated by an end-to-end text-to-speech (TTS) system based on an extended Tacotron architecture. The proposed TTS system includes encoders for speaker and emotion embeddings, a sequence-to-sequence text generator for creating Mel-spectrograms, and a WaveRNN to generate audio from the Mel-spectrograms. Extensive experiments show that the quality of the generated emotional speech can significantly improve SER performance on multiple datasets, as demonstrated by a higher mean opinion score (MOS) compared to the baseline. The generated samples were also effective at augmenting SER performance.
\end{abstract}


 \maketitle

\section{Introduction}
\label{sec:intro}

Speech emotion recognition (SER) is a rapidly growing field with many applications in fields such as healthcare, customer service, media, education, and forensics. While deep learning (DL) has shown promise in developing SER systems, their performance is still limited by the scarcity of emotion datasets \cite{latif2021survey,latif2022survey}. Existing SER corpora are small since the process of creating emotional data is costly and time-consuming, as multiple annotators have to manually listen to and annotate the material \citep{latif2022multitask,parthasarathy2020semi}. To increase data size, some studies have used multiple corpora, but the number of standard benchmark datasets is also limited, hindering progress in SER systems \citep{latif2020deep}.

Researchers have long been interested in creating natural-sounding TTS systems. TTS technology has come a long way from early TTS systems that often used pre-recorded waveforms pieced together based on input text \citep{hunt1996unit}. Such systems were prone to boundary artefact issues and statistical techniques were later developed to generate smoothed audio features for the vocoder to synthesise speech \citep{tokuda2000speech,zen2009statistical}. More recently, end-to-end neural network-based approaches have been proposed that can synthesise more natural-sounding human speech \citep{ren2019fastspeech,arik2017deep}. Current state-of-the-art TTS systems are trained using DL algorithms in an \textit{end-to-end} fashion, with popular models including Tacotron \citep{wang2017tacotron}, Deepvoice \citep{arik2017deep}, Fastspeech \citep{ren2019fastspeech,ren2020fastspeech}, Fastpitch \cite{lancucki2021fastpitch}, to name a few.

Unlike traditional systems, end-to-end TTS models can learn to generate a spectrogram directly from text without any complex pre-processing. These models, however, are currently only able to synthesise natural speech. Using generative DL techniques such as generative adversarial networks (GANs) \citep{goodfellow2014generative} for emotional speech synthesis is also challenging, as it requires a large amount of time-aligned data of a single speaker speaking the same content in different emotions and complex equations to guide the model in converting emotions using audio features. Some studies have achieved promising results in single-speaker emotional speech synthesis using TTS models \citep{kwon2019effective}, but the quality of synthetic speech in augmenting SER has not been evaluated.

In this paper, we propose a method for augmenting SER systems using an emotional text-to-speech (TTS) system and make two main contributions. \textit{Firstly}, we develop an \textit{end-to-end multi-speaker emotional} TTS system that does not require any alignment of audio files for emotion conversion or complex pre-processing of input data. Inspired by the success of end-to-end TTS models, we adopt a similar architecture to Tacotron. We propose to use a condition encoder to control the speakers' voices and emotions in the output speech. We generate speaker voice feature vectors using the encoder network. These feature vectors are modulated with one of the encoded emotional feature representations. These modulated feature vectors are used to condition the Tacotron to synthesise speech in different speaker voices and emotions. Subjective evaluation tasks show that our proposed model improves controllability and successfully synthesises emotional speech. \textit{Secondly}, we use the synthesised emotional speech to augment an SER system and conduct multiple experiments to evaluate the generated data quantitatively. Results show that the synthesised data can help improve SER performance in both within-corpus and cross-corpus settings.


The rest of the paper is organised as follows. In Section \ref{sec:format}, we briefly introduce the related work to change different features of audio. The model's architecture, loss functions, and flow of our architecture are described in Section \ref{sec:ar}. The details of the dataset and experimental condition in which we trained our model and hyper-parameters are provided in Section \ref{experiments}. We report our results in Section \ref{results}. Finally, this paper is concluded in Section \ref{conclusions}.

\section{Previous Work}
\label{sec:format}

In this section, we review the literature that has emerged around (1) the use of Tacotron for TTS, and for (2) emotional speech synthesis, and (3) the process of augmenting SER.

\subsection{Tacotron Based TTS Systems}

Many recent studies have focused on modifying the 
Tacotron model in order to better control the output of TTS systems. For instance, \cite{jia2018transfer} presented a Tacotron-based model that synthesises multi-speaker speech by conditioning the Tacotron on the speaker's voice embedding, which was generated from a speaker verification model \citep{wan2018generalized}. \cite{wang2018style} introduced a Tacotron variant that can change the speaking style, by learning different styles and saving them as vectors or tokens. These tokens are obtained by clustering similar accents and representing each cluster with an average. During synthesis, the Tacotron is conditioned on one of these tokens to produce speech with a specific style. \cite{skerry2018towards} presented a multi-speaker Tacotron that can change accents (e.g., American, Indian, British). Their model uses two encoder networks with the Tacotron and requires two audio samples (one for the accent and one for the speaker's voice) as input to generate the desired output. \cite{sun2020fully} proposed a Tacotron model that is trained with encoded output audio from a variational autoencoder as input. This not only improves the multi-speaker performance of Tacotron but also allows for control over the energy of the generated audio through the mean-variance property of the variational autoencoder. \cite{yasuda2019investigation} developed a Tacotron model that can learn more complex vocalisations by using the self-attention mechanism in Tacotron to learn complex dependencies related to pitch in different accents. They claim that their model outperforms traditional end-to-end approaches for languages with more pitch-dependent accents, such as Japanese. Our proposed model also generates speech in a multi-speaker setting and includes additional control over the emotions in the output.

\subsection{Tacotron Based Emotional TTS Systems}


Several previous works have attempted to generate emotional speech using TTS systems. For example, \cite{um2020emotional} developed an emotion control method for a TTS system based on the GST-Tacotron network \citep{skerry2018towards}, and demonstrated its effectiveness in synthesising emotional speech in a single-speaker setting in Korean. \cite{lee2017emotional} also evaluated a Tacotron-based emotional speech synthesizer in Korean, and found improvements in the quality of the generated speech for a single speaker. Other studies, such as~\cite{kwon2019effective,kim2020emotional}, have also proposed methods for controlling emotional speech synthesis, but these approaches only synthesise emotional speech in a single speaker's voice. In contrast, our proposed method achieves control over emotional speech synthesis for multi-speaker TTS and we also evaluate the quality of the synthesised data to augment the SER system.

\subsection{Augmenting Techniques for SER}

Speed perturbation~\citep{ko2015audio} is a popular data augmentation technique that has been widely studied in different contexts \citep{latif2019direct,aldeneh2017using}. It has been found to improve speech emotion recognition (SER) performance by creating copies of input data with different speed effects. Mixup~\citep{zhang2018mixup} is another data augmentation technique that generates augmented samples as a linear combination of original samples from the input data. Several studies have demonstrated the effectiveness of mixup in SER, including Latif et al.~\cite{latif2020deepar}, who used the technique to augment an SER system and achieve improved performance and robustness. A recent method called SpecAugment \citep{park2019specaugment}, originally proposed for automatic speech recognition, has also been applied to SER \citep{baird2021emotion}. In this study, the authors augmented the SER system with duplicate samples by a factor of two and found that SpecAugment improved model performance. Other studies \citep{latif2019direct,aldeneh2017using,latif2020federated} have also achieved improved performance by using input perturbation-based data augmentation techniques to increase the training data.

Further research is required to explore data-driven approaches  to increase the training data for SER. In this paper, we propose to explore TTS based data augmentation method where we explored different variations in the training data by changing the speaker and gender voices in different emotions. 

\section{Proposed Framework}
\label{sec:ar}
We propose to generate synthetic speech using a Tacotron-based emotional TTS system. We use synthetic speech data to augment the speech emotion classifier. The details of both emotional TTS and classifier are presented next. 

\subsection{Emotional Speech Synthesis}

\begin{figure*}
\centering
\includegraphics[width=0.8\textwidth]{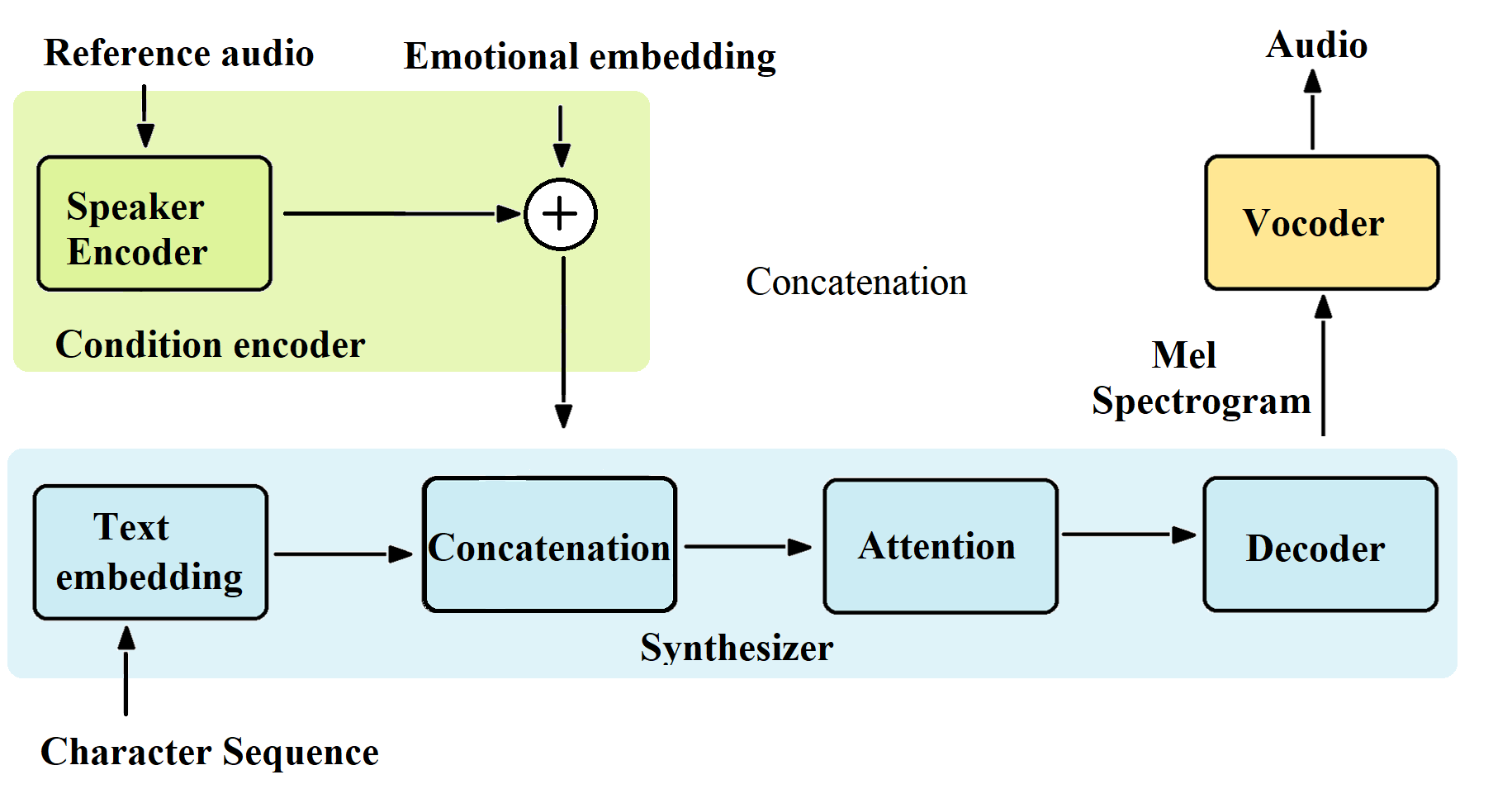}
\caption{Architectural flow diagram. The reference speaker's voice is first encoded and then modulated to desired emotion as described in the model schema. The output is then passed to the Tacotron decoder with the text embedding to synthesise the Mel-spectrogram.}
\label{fig:poModel}
\end{figure*}

Our model consists of an encoder which conditions Tacotron (as depicted in Figure \ref{fig:poModel}) to alter the speaker's voice and emotion in the output. Tacotron generates a Mel-spectrogram from a given text and embedding vector, while a Wave-RNN-based vocoder is used to generate an audio signal from the Mel-spectrogram

\subsubsection{Condition Encoder}

We propose using a condition encoder to create an embedding that represents both speaker identity and emotion. To do this, we use a speaker identification model presented in~\cite{wan2018generalized}, which creates a fixed-dimensional embedding, known as a d-vector \citep{variani2014deep,heigold2016end}, using a sequence of Mel-spectrograms computed from a speech signal of arbitrary length. We train this model using an end-to-end speaker verification loss that maximises the cosine similarity between utterances from the same speaker while minimising the cosine similarity between utterances from different speakers. We fine-tune this network on an emotional corpus to create an emotional embedding as well. Thus, the condition encoder is optimised to maximise the cosine similarity between embeddings of the same speaker with different emotions and to minimise the similarity between different emotions and different speakers. In this way, the model learns to generate a feature vector that contains both emotion and speaker identity information. The speaker's voice audio and emotion audio are embedded using the condition encoder and combined to generate a final embedding, which is used to condition the synthesizer to output speech with the selected emotion and speaker's voice.


For each unique emotion of every speaker in dataset, a centroid $c_{k}$ is calculated by taking the average of embedding for each unique emotion of every unique speaker. Loss for an embedding $e_i$ when the embedding and the centroid $c_{k}$ have the same speaker and emotion is calculated as:
\begin{equation}
\label{eq1}
    \mathcal{L}{(e_{i},c_{k})} = -1 \times \sigma({\cos(e_{i},c_{k})})
\end{equation}
When $e_{i}$ have different emotion or different speaker for centroid $c_{k}$ then loss is calculated as:-
\begin{equation}
\label{eq2}
    \mathcal{L}{(e_{i},c_{k})} = \sigma({\cos(e_{i},c_{k})})
\end{equation}
\begin{equation}
\label{eq3}
  {\mathcal{L}_{G}(S)} = \mathop{\sum}_{i,k} {L(e_{i},c_{k})}
\end{equation}

Equation \ref{eq1} maximises the cosine similarity between embeddings for the same speaker voice and same emotion. Equation \ref{eq2} represents the cosine similarity between embedding and centroid when they have different speaker voices or different emotions or both. Equation \ref{eq3} represents the final loss over every embedding, which is calculated as the sum of the loss for every embedding with every centroid.

The condition encoder consists of three LSTM layers with 768 cells each, and a final 256-length fully connected layer. The input to the model is the Mel-spectrogram generated from a speech utterance of a reference speaker's audio sample, and its output is an embedding vector of size 256 which represents the speaker's identity. After training the model, we use it to extract speaker and emotional information from a given audio. To separate the emotion from the speaker's voice, we generate vectors that only contain emotional information by using the trained condition encoder to generate embedding vectors for both the neutral and emotional voices of the same speaker. The neutral embedding vector is then subtracted from the emotional ones using Equation (\ref{eq4}), resulting in a vector that only contains emotional information. This vector can be used at inference time to control the emotion of the synthesised audio.
\begin{equation}
\label{eq4}
    {emb_{\text{em}}} = ({emb_{\text{en}}}-{emb_{\text{neu}}})
\end{equation}
Where $emb_{\text{en}}$ represents the embedding with emotion and voice information generated from the emotional voice of a speaker; $emb_{\text{neu}}$ is generated from neutral audio of the same speaker, and $emb_{\text{em}}$ represents the embedding that only contains emotional information. During inference, reference audio embedding (voice in which we want our output sample to be synthesised) and emotional embedding are added to generate a final embedding vector.
\begin{equation}
\label{eq5}
    {emb_{\text{final}}} = {emb_{\text{ref}}}+{emb_{\text{em}}}
\end{equation}
Finally, the modulated embedding vector and text are fed to Tacotron, which generates the Mel-spectrograms. These Mel-spectrograms are converted to the time domain using a vocoder, resulting in an audio signal.

\subsubsection{Synthesizer architecture}

The synthesizer is a variation of Tacotron \citep{wang2017tacotron}, which is a sequence-to-sequence model that generates output one frame at a time based on the input. In addition, we condition this synthesizer on an embedding vector generated by the condition encoder, which contains information about the desired output emotion and the speaker's voice. The condition embedding is concatenated with the text embedding of the synthesizer and then passed through a decoder to synthesise the output Mel-spectrogram. The synthesizer was trained on 80-channel Mel-spectrograms with a window size of 50 ms and a hop size of 12.5 ms. The synthesizer encodes the input characters into a hidden representation using three convolution layers, which learn longer-term context like an n-gram. The output of these convolution layers is passed to a single bi-directional LSTM layer with 256 units, which learns time dependencies from these n-gram-like features. The LSTM layer returns an encoded vector that fully represents the input text sequence. This vector is concatenated with a vector of emotional and speaker embeddings from the encoder.


It is worth noting that at this point, the encoder has been trained and its weights are not updated. The combined text, speaker, and emotion embedding is passed to the decoder to generate a Mel-spectrogram. The decoder architecture includes a location-sensitive attention mechanism that transforms the input embedding into a fixed-length vector. The output frame from the previous step is passed through two fully connected layers and concatenated with the embedding vector to ensure that sequences are generated without any time artefacts. This vector is then passed through two LSTM layers, and a linear transformation is applied to generate the next frame of the Mel-spectrogram. The output from this LSTM is also projected down to a single scalar, which serves as a stop token and indicates when to stop generating further frames. Once the Mel-spectrogram has been generated, it is passed through a 5-layer convolution network called the PostNet to improve overall reconstruction.


\subsubsection{Vocoder}

Traditionally, the Griffin-Lim algorithm \citep{griffin1984signal} was used to generate time-domain audio from a spectrogram, but it was slow and the output speech lacked naturalness. To address this, we use a vocoder based on the WaveRNN architecture \citep{kalchbrenner2018efficient}, which is a faster and more powerful recurrent network for sequential modelling of high-fidelity audio. It employs residual convolutions and GRU layers to generate a time-domain audio signal frame by frame from a Mel-spectrogram.

\subsection{Emotion Classifier}

To evaluate the synthesised emotions, we trained a deep neural network (DNN) for SER. We implemented a convolutional neural network (CNN)-based classifier that consists of a convolutional layer, a batch normalisation layer, and a dense layer before the softmax layer. Mel-frequency cepstral coefficients (MFCCs) are used as the input to the classifier. The CNN layers learn high-level features from the input features, which are then transformed by the dense layer into a more discriminative space for better emotion classification after passing through the normalisation layer.



\section{Experimental Protocol}
\label{experiments}
This section describes the details of the dataset, input feature, and model training. 

\subsection{Datasets}


We used the Librispeech dataset \citep{panayotov2015librispeech} to train our TTS model. It consists of 1000 hours of speech data from various speakers, sampled at 16 kHz. For the emotion embeddings, we used the Emotional Voices Database (EVD) \citep{adigwe2018emotional} and the Toronto Emotional Speech Set (TESS) \citep{dupuis2010toronto}, which contain six different speakers reading different sentences with different emotions. We conducted multiple experiments to evaluate the performance of our model. For emotion classification experiments, we used the Ryerson Audio-Visual Database of Emotional Speech and Song (RAVDESS) \citep{livingstone2018ryerson} and TESS. For cross-corpus emotion classification, we used the CREMA-D \citep{cao2014crema}, SAVEE \citep{jackson2014surrey}, EmoDB \citep{burkhardt2005database}, and synthesised audio. The details of these datasets are presented in Table \ref{data}. We used one speaker from Librispeech, as well as all the speakers from EVD and TESS with two samples that were not included in the training set, to determine the mean opinion score. For emotion classification experiments, we use speaker-independent emotion classification. We randomly select 70\% of CREMA-D for training, 10\% for validation, and 20\% for testing. The full corpora including RAVDESS and EmoDB were used as the test set in the emotion classification experiments, and the SAVEE dataset was used as the test set in the cross-corpus emotion classification experiments.


\subsection{Input Features}


Tacotron takes text strings as input, which are sequences of characters. Each character is encoded into a one-hot encoded vector and embedded in a continuous vector. The other input to Tacotron is a condition embedding vector that contains speaker and emotion information. This vector is obtained from an encoder, which takes speaker audio as input and converts it into Mel-frequency cepstral coefficients (MFCCs). These MFCCs have 40 log filter banks, 80 frames, and no overlapping window. To generate t-distributed stochastic neighbour embedding (t-SNE) plots of synthesised audio, we encoded our synthesised audio using the model presented in~\cite{jia2018transfer}. The input to this model is also MFCCs with 40 log filter banks, 80 frames, and no overlapping window, resulting in an 80x40-dimensional feature vector. This model is also used in evaluating the equal error rate (EER) in speaker verification. In emotion classification and cross-corpus emotion classification, we use MFCCs with 40 log filter banks and a hop size of 64 milliseconds. The MFCC array is transposed and the arithmetic mean is calculated across its horizontal axis as in a previous work \citep{de2020emotions}.



\begin{table}[!ht]
\caption{Description of all the considered datasets.}
\centering
\begin{tabular}{|l|c|c|} 
\hline
Name                                                                   & \begin{tabular}[c]{@{}l@{}}Number of\\Speakers\end{tabular} & \begin{tabular}[c]{@{}l@{}}Number of\\Utterances\end{tabular}  \\ 

\hline
CREMA-D                                                                & \textcolor[rgb]{0.125,0.125,0.125}{91}                      & 7,442                                                                                                                 \\ 
\hline
EmoDB                                                                 & 10                                                          & 535                                                                                                                  \\
\hline
\begin{tabular}[c]{@{}l@{}}EVD\end{tabular} & 5                                                           & 7,590                                                                                                                 \\ 

\hline
Librispeech                                                            & 2484                                                        & 281,241                                         \\ 
\hline
REVDESS                                                                & \textcolor[rgb]{0.125,0.125,0.125}{24~}                     & 7,356                                                                                                                 \\ 

\hline
SAVEE                                                                  & 4                                                           & 480               
\\ 

\hline
TESS                                                                   & 2                                                           & 2,800                                                                                                                 \\ 

\hline
\end{tabular}
\label{data}
\end{table}

\subsection{Speech Synthesis Models Training}

First, the encoder is trained on the Librispeech dataset to learn to generate a speaker embedding that is distinct for each speaker. It takes a Mel-spectrogram as input and outputs an embedding vector of size 256. From these embedding vectors, a similarity matrix is constructed such that each column contains an embedding vector for a unique speaker, and cosine similarity is maximised in all cells of the columns and minimised in all cells of the rows. Cosine similarity is maximised along the columns because they contain audio embeddings for the same person, whereas it is minimised along the rows because they contain audio embeddings for different people. In this way, the embeddings of the same people are similar and those of different people are different.


After training the encoder on the Librispeech data, it is fine-tuned on the EVD and TESS datasets to generate distinct embedding vectors for different emotions. This time, a similarity matrix is constructed such that a column contains embedding vectors generated for a single emotion for the same speaker, and other emotions are placed in other columns. This is done for all speakers, and then cosine similarity is maximised along a column and minimised across columns. This is done to increase the distance between different emotions of the same person, so cosine similarity is minimised by adding it across columns rather than within the same columns. We used a batch size of 30 and a learning rate of $10^{-4}$. 

During training, the synthesizer model is first trained on the Librispeech data so that it can learn to generate audio of different speakers from a diverse range of text. This is because the EVD and TESS datasets combined only have six speakers. Once the synthesizer is trained enough that it can generate audio resembling the reference speaker, we fine-tune it to generate different emotional Mel-spectrograms by training it on the EVD and TESS datasets. We use a learning rate of $10^3$ that exponentially decays to $10^{-5}$, and a batch size of 30 for training the synthesizer. The Adam optimiser with $\beta_1 = 0.9$, $\beta_2 = 0.999$, and $\epsilon = 10^{-6}$ is used as the optimiser. The teacher forcing ratio is set to 1 (meaning the original previous sequence is shown to the model for prediction of the next sequence). The mean squared error is minimised for the predicted Mel-spectrogram.

\section{Results}
\label{results}

In this section, we evaluate the performance of our proposed model in terms of the similarity of the synthesized speakers and the granularity of synthesized emotions.

\subsection{Evaluating Synthetic Speech Quality}
To evaluate the quality of synthetic speech, we conducted multiple experiments. The details of these experiments are presented below.

\subsubsection{Speaker Verification}


We evaluated the speaker similarity of synthesised audios with real speech using speaker verification and measured the equal error rate (EER) following \cite{jia2018transfer}. The EER is used to measure the performance of a speaker verification system by comparing the false reject rate (FRR) and false accept rate (FAR) at different sensitivity levels. The EER is the point at which the FRR and FAR are equal. To calculate the EER, we used 100 audio samples, 40 of which were synthesised. We enrolled only synthesised speakers in the system and calculated the EER. We achieved an EER of 0.10\% by performing voice conversion using a multi-speaker Tacotron model \citep{jia2018transfer}. We also generated emotional audio samples using a base model, and the speaker verification model gave an EER of 0.24\% on these synthesised audios. In contrast, we achieved an EER of 0.16\% when using the proposed model for both emotion and voice conversion. The EER on real samples using the approach in \cite{jia2018transfer} was 0.04\%. We have compared the EER of these models in Table \ref{EER}.

\begin{table}[!ht]
\caption{Speaker verification EERs of different synthesizers.}
\centering
\begin{tabular}{|l|c|l|}
\hline
              & \# of samples & EER  \\ \hline
Emotion + voice conversion TTS & 100           & \textbf{0.16} \\ \hline
Baseline Emotion conversion TTS & 100           & 0.24 \\ \hline
Voice conversion TTS & 100           & 0.10 \\ \hline
Real audios & 100           & 0.04 \\ \hline
\end{tabular}
\label{EER}
\end{table}

\subsubsection{Listening Experiments}


We performed mean opinion score (MOS) evaluations to measure the quality of synthesised speech. We asked subjects with post-graduate exposure to give a score after listening to the audio based on the following standard: 1 = Bad; 2 = Poor; 3 = Fair; 4 = Good; and 5 = Excellent. The results, shown in Table \ref{mos}, indicate that the proposed model can synthesise high-quality emotional speech compared to the baseline model. The proposed model significantly improves the MOS score for emotions including angry, sad, and happy compared to the baseline. However, it achieves slightly lower MOS scores for natural speech compared to the baseline. This may be because the baseline model is specifically designed to generate natural speech and therefore performs better for neutral speech. Nevertheless, our proposed model performs well for all emotions. Readers can listen to samples of the generated speech at this URL\footnote{\url{https://emotaco.github.io/Emotional_Tacotron/}}.


\begin{table}
\captionsetup{justification=centering}
\caption{Mean Opinion Score (MOS) with 95\% confidence interval.}
\centering

\begin{tabular}{|l|l|l|l|l|l|}
\hline
Emotion & Angry & Happy & Sad  & Neutral & Overall \\ \hline
Recorded     & 4.6  & 4.50  & 4.50 & 4.60    & 4.55    \\ \hline
Baseline    & 2.80  & 3.10  & 2.70 & 4.20   & 3.20   \\ \hline
Proposed  & 3.60  & 3.70  & 3.80 & 4.10    & 3.80    \\ \hline

\end{tabular}
\label{mos}
\end{table}

\subsubsection{Speaker and Emotion Visualisation}

\begin{figure*}
\centering
\includegraphics[width=1\textwidth]{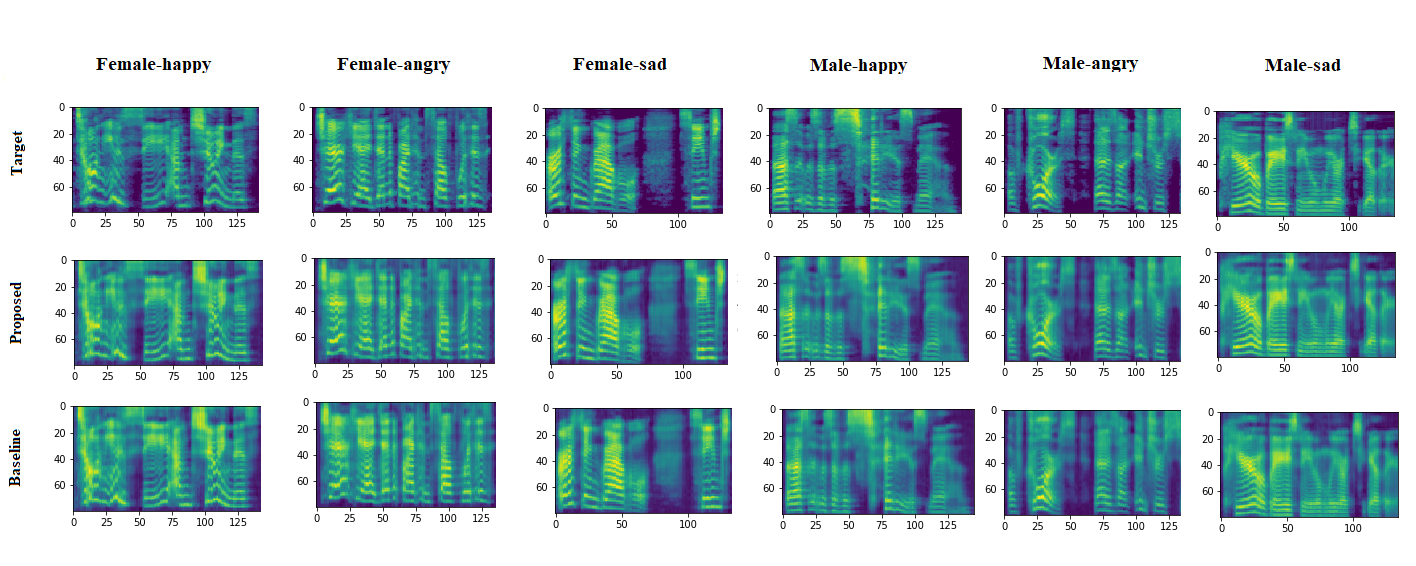}
\caption{Comparison of target and synthesized Mel-spectrograms for various emotions in Male and Female audios.}
\label{fig:spec}
\end{figure*}

During this experiment, we did not use teacher forcing and generated audio as described in the inference part. The synthesised Mel-spectrograms for different emotions by the baseline and proposed models were plotted in Figure \ref{fig:spec}, and the results were compared with the target Mel-spectrograms. In contrast to the baseline, our proposed model did not smooth the generated Mel-spectrograms that help produce a better quality of emotional speech using WaveRNN vocoder.

\begin{figure}
\centering
\includegraphics[width=0.5\textwidth]{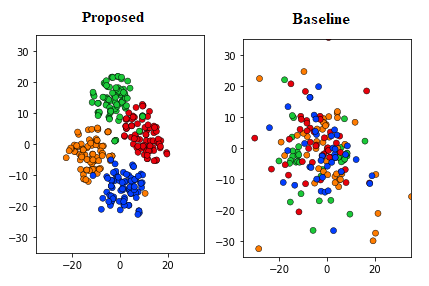}
\caption{Comparison of t-SNE plots of male audio for various emotions using baseline and our proposed model shows that our model demonstrates better emotion performance.}
\label{fig:maleTSNE}
\end{figure}

\begin{figure}
\centering
\includegraphics[width=0.5\textwidth]{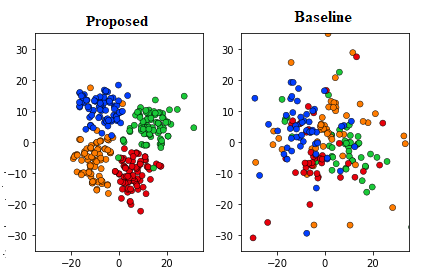}
\caption{Comparison of t-SNE plots of female audio for various emotions using baseline and our proposed model shows that our model demonstrates better emotion performance.}
\label{fig:femaleTSNE}
\end{figure}

For the purpose of evaluation, we present the t-SNE plot, which was generated by embedding vectors generated from synthesised output samples using a speaker verification model as the encoder. Note that the speaker encoder was not trained with the synthesizer, so it is not optimised for synthesizer output. We generated t-SNE plots for emotional audio synthesised using the model from the base papers and compared the results with the proposed model. These t-SNE plots for synthesised speech in both male and female voices are shown in Figure \ref{fig:maleTSNE} and \ref{fig:femaleTSNE}, respectively. These plots demonstrate that our model is able to synthesise distinct emotions compared to the base model. It can be observed that different emotions are separated and similar emotions are clustered together, indicating similarity between emotions. 


\begin{figure}[!ht]
\centering
\includegraphics[width=0.5\textwidth]{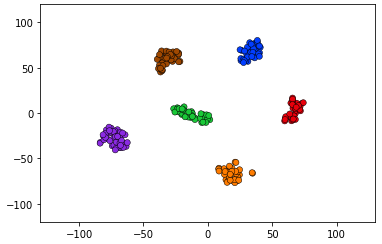}
\caption{The t-SNE plot for speaker voice of synthesised results shows that individual speakers' voices are distinctly clustered together.}
\label{fig:speakerTSNE}
\end{figure}

Since the angry emotion has more expression compared to the sad and happy emotions, which are tone variations, the cluster of angry emotions is farther from the happy emotions. We also visualise the t-SNE plot of multiple speakers in neutral speech using our proposed model in Figure \ref{fig:speakerTSNE}. It shows distinct clusters for different speakers indicating that the model is able to learn the multiple speaker embeddings effectively. 

.

\subsection{Augmenting Speech Emotion Recognition (SER)}
In this section, we used the synthetic speech to augment the SER system. We performed our evaluations using corpus and cross-corpus settings. Results for these experiments are presented next. 

\subsubsection{Within Corpus Evaluations}

\begin{figure}[!ht]
\centering
\includegraphics[width=0.4\textwidth]{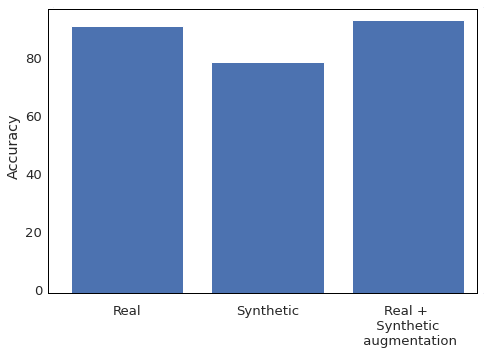}
\caption{Bar plot which shows that our synthesized audio's emotion and real audio emotions are almost similarly classified by the classification model. }
\label{fig:bar}
\end{figure}

 We used the RAVDESS and TESS datasets for evaluations. We combined both datasets and then randomly split the data into a ratio of 70:10:20 for train, validation, and test sets, respectively. We trained the model for 45 epochs. We compared the results for speaker recognition on real and synthesised speech in Figure \ref{fig:bar}. We achieved an accuracy of 80\% for synthesised speech, while the accuracy for the real speech test set was 92.4\%. This demonstrates that our model can synthesise the emotional characteristics of output speech. We also augmented the classifier with synthetic data and performed training using both real and synthesised speech data. We achieved an accuracy of 94.6\%, which is better compared to the classifier trained on real data alone. This experiment shows that our model can also be used to generate additional audio data which can be used to augment speaker recognition systems to improve their performance.

\begin{figure*}
\centering
\captionsetup{justification=centering}
\includegraphics[width=1\textwidth]{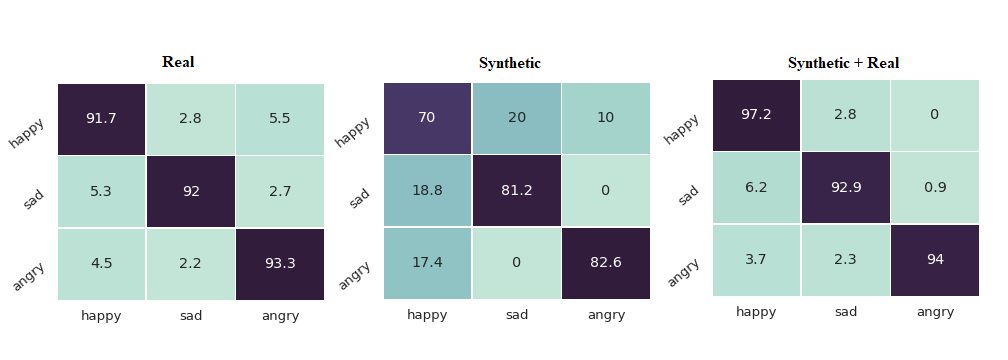}
\caption{Confusion matrix for the test set of real, synthetic, and combined synthetic and real audio. The addition of synthetic data improves emotion classification.}
\label{fig:conf}
\end{figure*}

We have also plotted confusion matrices in Figure \ref{fig:conf} for emotion classification on real audio, synthetic audio, and a combination of real and synthetic data in the training set. The confusion matrix shows that the model augmented with synthetic data is able to better classify speech emotions. The accuracy of other emotions has also been improved, but the most significant improvement can be seen in the classification of happy emotions.

\subsubsection{Cross-Corpus Corpus Evaluations}

\begin{figure}[!h]
\centering
\includegraphics[width=0.5\textwidth]{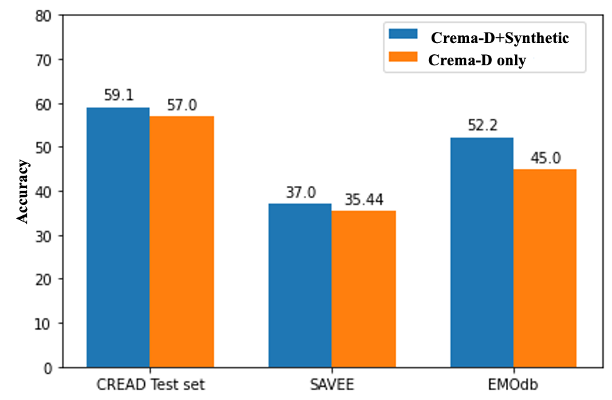}
\caption{Test results in cross-corpus setting, which shows improvements when the model is augmented with synthetic data. }
\label{fig:cross}
\end{figure}

We also evaluated the effect of augmenting with synthetic data by performing cross-corpus emotion classification. To do this, we implemented a classifier consisting of an LSTM layer, three dense layers, and a softmax layer for emotion classification. We also used two dropout layers between dense layers to learn more generalised representations. We selected the architecture of the model based on previous research findings \citep{latif2019unsupervised,latif2019direct}. We trained the classifier on MFCC features extracted from the input audio. The model was trained with a sparse categorical cross-entropy loss and Adam optimiser for 100 epochs. The model was trained using the CREMA-D dataset and the CREMA-D dataset augmented with synthetic data and was evaluated on the CREMA-D, SAVEE, and EMODB datasets. The results, shown in Figure \ref{fig:cross}, demonstrate that adding synthesised data increases accuracy not only on the SAVEE and EMODB datasets without fine-tuning the model but also on the CREMA-D test set as well.

\subsubsection{Changing Gender and Speaker Distributions}


In this experiment, we compare the results of data augmentation with new speaker voices that are not present in the given corpus. For instance, the SAVEE corpus has four male speakers, and synthetic data can be created either in the voices of these four male speakers or in the voices of additional male and female speakers to bring diversity to the data and augment speech emotion classification. We present the results in Table \ref{speaker_Dis}. We compared the results with the baseline model, which was trained without any augmentation, and also with the application of speed perturbation to the training data. We followed \citep{latif2020deep} and created two copies of augmented samples using the speed perturbation data augmentation technique. We found that augmenting the data with different speaker voices helps improve performance compared to the baseline and the widely used data augmentation technique of speed perturbation.

\begin{table*}[!h]
\centering
\caption{Results using different distributions of synthetic data for speakers and gender}
\begin{tabular}{|l|lllll|}
\hline
\multirow{2}{*}{Dataset} & \multicolumn{5}{c|}{Accuracy (\%)}                                                      \\ \cline{2-6} 
                         & \multicolumn{1}{l|}{Baseline} & \multicolumn{1}{l|}{\begin{tabular}[c]{@{}l@{}}Speed perturbation\\ augmentation\end{tabular}} & \multicolumn{1}{l|}{\begin{tabular}[c]{@{}l@{}}Male spakers\\ synthetic data\end{tabular}} & \multicolumn{1}{l|}{\begin{tabular}[c]{@{}l@{}}Female speakers\\ synthetic data\end{tabular}} & \begin{tabular}[c]{@{}l@{}}Both female and \\ male synthetic data\end{tabular} \\ \hline
SAVEE    &   65.4    & \multicolumn{1}{l|}{66.8}    & \multicolumn{1}{l|}{68.2}    & \multicolumn{1}{l|}{69.4}      &    72.3                  \\ \hline
CREMA-D     &  68.3    & \multicolumn{1}{l|}{70.1} &  \multicolumn{1}{l|}{72.7}             & \multicolumn{1}{l|}{72.9}                                                                         &  74.3      \\ \hline   
\end{tabular}
\label{speaker_Dis}
\end{table*}

\section{Conclusions}
\label{conclusions}


This paper proposes to utilise an emotional text-to-speech (TTS) system to augment a speech emotion recognition (SER) system. We present a Tacotron-based multi-speaker emotional TTS system for synthetic speech generation in different speaker voices and use it for data augmentation in speech emotion recognition to improve performance. The results showed that the proposed TTS system can generate high-quality emotionally discriminative samples. When we augment the SER system with these augmented samples, we find that using synthetic data in different emotional voices can help improve performance compared to the widely used speech data augmentation technique in SER. Our future work will focus on investigating the learning of a unified embedding for controlling style and emotions for all people, regardless of age, background, and gender.



\end{document}